\documentclass[aps,prd,10pt,twocolumn,showpacs,preprintnumbers,amsmath,amssymb,nofootinbib]{revtex4-1}

\usepackage{natbib} %Default!
\usepackage{color}
\usepackage{amsmath}
\usepackage{epsfig}
\usepackage{array}
\usepackage{graphicx}

\begin{document}

\title{A Game-Centered, Interactive Approach for Using Programming Exercises in Introductory Physics}

\author{Orban, C.$^*$}
\affiliation{Department of Physics, The Ohio State University, Columbus, OH, 43210}
\author{Porter, C. D}
\affiliation{Department of Physics, The Ohio State University, Columbus, OH, 43210}
\author{Smith, J. R. H.}
\affiliation{Department of Physics, The Ohio State University, Columbus, OH, 43210}
\author{Brecht, N. K.}
\affiliation{Department of Physics, The Ohio State University, Columbus, OH, 43210}
\author{Britt, C. A.}
\affiliation{ASC Tech, The Ohio State University, Columbus, OH, 43210}
\author{Teeling-Smith, R. M.}
\affiliation{Marion Technical College, Marion, OH, 43302}
\affiliation{University of Mt. Union, Alliance, OH, 44601}
\author{Harper, K. A.}
\affiliation{Department of Engineering Education, The Ohio State University, Columbus, OH, 43210}

\email{orban@physics.osu.edu}

\date{\today}

\begin{abstract}

Incorporating computer programming exercises in introductory physics is a delicate task that involves a number of choices that may have a strong affect on student learning. We present an approach that speaks to a number of common concerns that arise when using programming exercises in introductory physics classes where most students are absolute beginner programmers. These students need an approach that is (1) simple, involving 75 or fewer lines of well-commented code, (2) easy to use, with browser-based coding tools, (3) interactive, with a high frame rate to give a video-game like feel, (4) step-by-step with the ability to interact with intermediate stages of the ``correct" program and (5) thoughtfully integrated into the physics curriculum, for example, by illustrating velocity and acceleration vectors throughout. We present a set of hour-long activities for classical mechanics that resemble well-known games such as ``asteroids", ``lunar lander" and ``angry birds". Survey results from the first activity from four semesters of introductory physics classes at OSU in which a high percentage of the students are weak or absolute beginner programmers seems to confirm that the level of difficulty is appropriate for this level and that the students enjoy the activity.  These exercises are available for general use at \url{http://compadre.org/PICUP} In the future we plan to assess conceptual knowledge using an animated version of the Force Concept Inventory originally developed by \cite{Dancy2006}.

%Incorporating computer programming exercises in introductory physics is a delicate task that involves a number of choices that may have a strong affect on student learning. We present an approach that speaks to a number of common concerns that arise when using programming exercises in introductory physics classes where most students are absolute beginner programmers. These students need an approach that is (1) simple, involving 75 lines of well-commented code or substantially fewer, (2) easy to use, with browser-based coding tools, (3) interactive, with a high frame rate to give a video-game like feel, (4) step-by-step with the ability to interact with intermediate stages of the ``correct" program and (5) thoughtfully integrated into the physics curriculum, for example, by illustrating velocity and acceleration vectors throughout. We present a set of hour-long activities for classical mechanics that resemble well-known games such as ``asteroids", ``lunar lander" and ``angry birds". Importantly, these activities take advantage of the game-like environment to give students a feel for the physics. We plan to assess learning gains using an animated version of the Force Concept Inventory originally developed by \cite{Dancy2006}. We argue that the physics education research community needs to direct more attention to answering basic questions that we outline here about the best way to incorporate programming exercises into introductory physics. These exercises are available for general use at \url{http://compadre.org/PICUP}
\end{abstract}

\maketitle

\section{Introduction}

The need to incorporate programming content into introductory physics is widely appreciated by the academic community \cite{Fuller2006}. By some estimates, \emph{at least 70\%} of new STEM jobs in the US will require computer programming skills \cite{labor2014} and in the sciences computer programming skills have become an essential part of many disciplines. In response to these shifts, groups like code.org and the ``hour of code" have brought coding tutorials to wider and younger audiences \cite{code}. These groups also influenced federal education legislation in the US. In particular, the Every Student Succeeds Act (ESSA), which was signed into law in December 2015, designates computer science as a ``core subject" for the first time. This is a significant change that places computer science on the same level as english and mathematics \cite{coresubject}. The 2017-2018 school year will be the first school year that this legislation will be fully implemented.  Yet, for physics instruction, and perhaps even more generally, the task of re-imagining STEM courses with computer science as a crucial element is still far from complete. Although there is significant research \cite{Caballero_etal2012} and a number of universities using vpython \cite{Chabay_Sherwood2008} in calculus-based introductory physics, vpython exercises are much less often used in algebra-based physics and at the high school level. There is some evidence that the approach is too difficult for this level. \cite{Aiken_etal2013} describes a masters degree project by a high-school physics teacher who worked for two years to develop a vpython curriculum for a 9th grade high school physics class and found that only 1/3$^{\textnormal{rd}}$ of the class successfully completed the exercises and perhaps only 20\% understood the iterative nature of the physics as intended. While \cite{Aiken_etal2013} used a small sample ($N = 32$), the results underscore the need to develop a curriculum that adds programming into \emph{algebra-based} physics with a higher success rate. In this paper we discuss an approach that that differs from the vpython exercises described by \cite{Aiken_etal2013} and in the vpython-based Matter \& Interactions curriculum \cite{Chabay_Sherwood2015} in important ways. The content we describe here also compliments the existing computational exercises that are hosted on the AAPT Partnership for Integrating Computation into Undergraduate Physics (compadre.org/PICUP) by providing content that is aimed at algebra-based physics and non-major physics courses \cite{PICUP}. For convenience we have made the exercises described here available on the PICUP site.

%Porter: I happen to agree with the "third canvas" comment below. But do you have any references that support that learning through programming reinforces/builds physical understanding in different ways than labs/lecture? The statement seems to put programming on the same level as lecture and lab, and (especially given what we know happens in lab) I tend to agree; it's just a very bold statement to make without backup. If we don't, then we might want to say something like "it is important to investigate the effect size of programming labs, because it may end up that it is as important as lecture or lab"... much more eloquent of course.
%CMO -- we may be able to find something written but Rubin Landau in the introduction of his computational physics books that talks about computational physics and "numerical experiments". The idea of "numerical experiments" is really key to the 3rd canvas metaphor. I imagine there would be things to cite from the STEM+C solicitation where they wax philosophical about the virtues of "computational thinking". Totally agree that we need many more references in this section.

What the approach we describe here and other computational physics education efforts have in common is a desire to change how students experience physics courses and how they think about physics \cite{Chabay_Sherwood2008,Sherin2001}. Programming activities represent a kind of third ``canvas" for illustrating physics ideas besides pen \& paper calculations and laboratory exercises \cite{Serbanescu_etal2011}. Whereas pen and paper activities tend to emphasize mathematical calculations that describe how forces take an initial state of a system to a final state over some time interval, programming activities emphasize how forces can act in an iterative sense over a number of much smaller time intervals to produce the same behavior. This may seem like a subtle distinction, but to a student learning physics for the first time, and who may have been exposed to computer programming ideas through activities like code.org, they may find programming activities in physics to be more intuitive than pen \& paper calculations or laboratory exercises and some authors argue that numerical exercises highlight  the physics of complicated interactions in a clearer way \cite{Buffler_etal2008,Chabay_Sherwood2008}. To other students who may be comfortable with pen \& paper calculations, programming activities provide another perhaps unfamiliar context to apply their physics knowledge. To the extent that expertise involves the ability to accurately apply concepts in unfamiliar situations \cite{Holyoak_etal1991}, programming activities can provide an interesting challenge to students -- do they understand physics well enough to program a computer to simulate a physics problem?

While there are plenty of reasons to be optimistic about adding programming content to introductory physics \cite[e.g.][]{muppet1993,Landau2006,Landau_etal2011,Weintrop_etal2016}, and in this paper we describe a specific implementation in use at Ohio State University (OSU) that we are beginning to test in high-school physics classes, it should be said that there are unanswered questions about whether these activities provide physics learning gains beyond conventional web interactives \cite[e.g.][]{Perkins_etal2006,Podolefsky_etal2010} where the student does \emph{not} see the underlying code. Another re-statement of this question is: under what conditions does working with a physics simulation code cause students to think more deeply about the physics content? As the title of this paper suggests, our hypothesis is that students must both develop a computer program step-by-step and participate in goal-oriented game-like activities in interacting with this program to appreciate the physics on a deeper level. Although there are some exceptions, goal-oriented game-like activities are typically not a part of programming exercises at the introductory level. In the Matter \& Interactions curriculum that integrates vpython into calculus-based physics \cite{Chabay_Sherwood2015}, many of these programs, such as a three-body gravitational simulation or the 3D pendulum \cite{Chabay_Sherwood2008}, are written for the student to passively watch except perhaps for changing the perspective. And while there are a large number of exercises currently available on the AAPT's Partnership for the Integration of Computation into Undergraduate Physics (compadre.org/PICUP), only few of them involve a high level of interactivity. 
%The same can also be said of computational physics textbooks that feature dozens of example programs \cite[e.g.][]{Landau_etal2015}. 
%While some vypthon programs do involve interactivity, such as the program where students use the mouse to draw the force vector and watch an object respond accordingly \cite{Chabay_Sherwood2015}, but these programs tend not have a specific goal or game-like quality and students may find little reason to play around with the animation for as long as it may take to internalize the physics that is being demonstrated. 
But there some groups that focus on science-based computer activities in primary and secondary schools appreciate the importance of video-game like goals \cite{Taub_etal2015,Weintrop_etal2016}.

Proving the value of game-like activities in programming exercises in a definitive way is beyond the scope of this paper. Instead we will here describe a set of seven different computer programming activities designed for absolute beginner programmers in introductory physics (mechanics) classes. Survey results from four semesters of student data that probe student experiences with the first exercise will be presented.
We also describe an assessment framework that we plan to use that is based on an animated version of the Force Concept Inventory \cite{Dancy2006} that we have reconstructed from various sources and in a way that that can be used on modern computers and without any java dependencies. A larger study with more definitive conclusions will be described in future work.

We believe that the exercises described in the next section fulfill many of the basic requirements of any approach to incorporating programming into introductory physics in classes with absolute beginner programmers. These students need an approach that is
\begin{enumerate}
    \item simple, involving about 75 or fewer lines of well-commented code, and no calculus knowledge
    \item easy to use, with browser-based coding tools that include real-time bug-finding capabilities
    \item interactive, with a fast frame rate ($\approx$60 frames per second), high-quality 2D graphics and responsive keyboard controls
    \item step-by-step, including the capability for students to interact with correct versions of the intermediate stages of the program but without seeing the correct code,
    \item and thoughtfully incorporated into the physics curriculum.
\end{enumerate}
%Porter: These are really fantastic. Where do they come from? What supports using these as criteria? Even though this is a preliminary statement of purpose paper, I think you might justify these criteria by referring to notes/observations made by two of the authors in teaching computational techniques to physics majors at OSU. I think we can justly say "teaching" since I taught it solo one year. 
%CMO -- I came up with this list myself over time. Part of it is opinions I formed in surveying things like VisualPython in summer 2014. I decided to use processing at that point. In fall 2014, used processing.js and I did not have any bug finding capabilities and this turned out to be a nightmare for me trying to debug every student's code. I also didn't have vectors illusrated then. In spring 2015, I added vectors and we started using the processing IDE with its great bug finding capabilities, but getting it installed on everyone's laptops turned out to be a real pain. Chris Britt has been working since June 2016 to enable in-browser coding.

The following sections will focus especially on the first and fifth items above. Regarding items 2, 3, and 4 we use a programming framework called \emph{processing} that was developed to give artists a simple but graphically powerful programming language to create interactive art \cite{processing}. We find that this framework, which uses a syntax similar to C/C++, is very capable of producing 2D physics interactives that run at a very high frame rate ($\approx 60$ frames per second) on modest computers, such as the chromebook laptops that most high schools have on hand. Frame-rate is key to producing smooth-running interactives that are passable as video games. \emph{Processing} also has built-in aliasing that takes advantage of high-resolution screens. Importantly, \emph{processing} can produce these high-quality interactive visualizations using any browser and without any required installations or plugins. This removes the tedious step of conferring with IT staff to install software on each computer and periodically upgrading this software if there are updates. The framework can also be used to allow students to interact with as many intermediate stages of the program as necessary without revealing the code. \emph{Processing} is a free resource with good documentation and an active, growing community. \emph{Processing} has two different browser-based coding methods, of which p5.js \cite{p5js} is the more developed method that we highlight here. 

Although there is good work in the literature describing how numerical exercises can be connected with laboratory exercises \cite[e.g.][]{Serbanescu_etal2011}, we consider this out of scope for the present work. \emph{Processing} does have capabilities to interact with Arduino circuit boards \cite{arduino,processing}, making this an interesting possibility for future work.

\section{Overview of Programming Activities for Mechanics}

In a semester course of introductory physics at Ohio State University at the regional campus in Marion, we include six required programming activities and a seventh activity that is optional or extra credit. In most other ways the course is identical to the same course on the Columbus campus. The official description of this course is calculus-based physics I, but on all OSU campuses students only need to be concurrently enrolled in calculus in order to take the course, and as a result the calculus content in the course is limited. Moreover, the students at OSU's regional campuses are less prepared than their peers on OSU's Columbus campus. Both of these facts make this course an interesting venue for integrating programming exercises into introductory physics with the end goal of creating a curriculum that might succeed in the high school physics classroom.

%Each activity is designed to take about an hour and for this reason we sometimes refer to these activities as programming ``labs", which is also an oblique reference to the idea that these exercises are also interactive numerical experiments. 
%In a typical course meeting, students are required to complete a conventional pen and paper calculation activity. Most students will be able to complete this activity with a half-hour remaining, at which point they are encouraged to work on programming activities until the end of the period. 
%These exercises have been in use in some form at OSU's regional campus in Marion since fall 2014. Regional campuses are ideal places to experiment with new course content before scaling up to much larger classes on the main campus. 

Each activity is designed to take about an hour to complete. To date, about 125 students from OSU's regional campus in Marion have completed the exercises described here. 

All of the exercises illustrate the velocity, acceleration and force vectors using the same color scheme for these vectors used by the textbook. All of the exercises build off of each other in a way that would make it hard for a student to start in the middle, and the first exercise gives the student much of the code that they will need, only asking them to modify certain parts of the code in certain ways as will be discussed in the next section. All of the exercises contain optional ``challenges" that encourage the student to develop some functionality that often adds an interesting element to the game. The last two exercises utilize a graphing system that we developed in order to show, for example, how $v_y$ changes with time in a projectile problem, or how the $x$ position of some object oscillates. The list of exercises is as follows:
\begin{enumerate}
    \item Planetoids (similar to the classic game ``Asteroids")
    \item Lunar descent (similar to the classic game ``Lunar lander")
    \item Bellicose birds (similar to the popular game ``angry birds")
    \item Planetoids with momentum
    \item Planetoids with torque
    \item Planetoids with a spring (harmonic motion)
    \item Extra credit: Bellicose birds with energy
\end{enumerate}

The choice of an asteroids-like game for the first activity is intentional. This puts the student in the frame of reference as a rocket in free space, which is a situation where the laws of physics are particularly simple. In the second exercise (``Lunar descent"), adding two lines of code to the planetoids exercise is all that is needed to add gravity to the interactive simulation. And in the third exercise (``Bellicose birds"), only a handful of changes are needed to convert the ``lunar descent" activity into something that resembles the popular game ``angry birds". The task in this exercise is to draw the expected trajectory of a bird-like projectile before it is launched. Exercises 4-7 will be discussed later.

This sequence is designed to accompany a typical physics course where momentum is not introduced until mid-way through the course, followed by concepts of torque and, later, harmonic motion. The energy exercise is made available to students in the middle of the course when energy is introduced, but this exercise is more difficult than the others as will be discussed later.

\section{Planetoids: The First Programming Activity}

The natural environment for illustrating Newton's laws is free space, away from any sources of gravity. In such an environment, objects in motion will of course continue with the same velocity, moving in a straight line, unless a force is acting on it. The classic game ``asteroids" illustrates this well with a ship that drifts through free space, except when its rockets fire to avoid asteroids that are also drifting through free space. The net force is either zero, or constant and in the direction the ship is pointing.

\subsection{Structure of the Program and Choices}

\begin{figure*}
    \centering
    \includegraphics[width=7in]{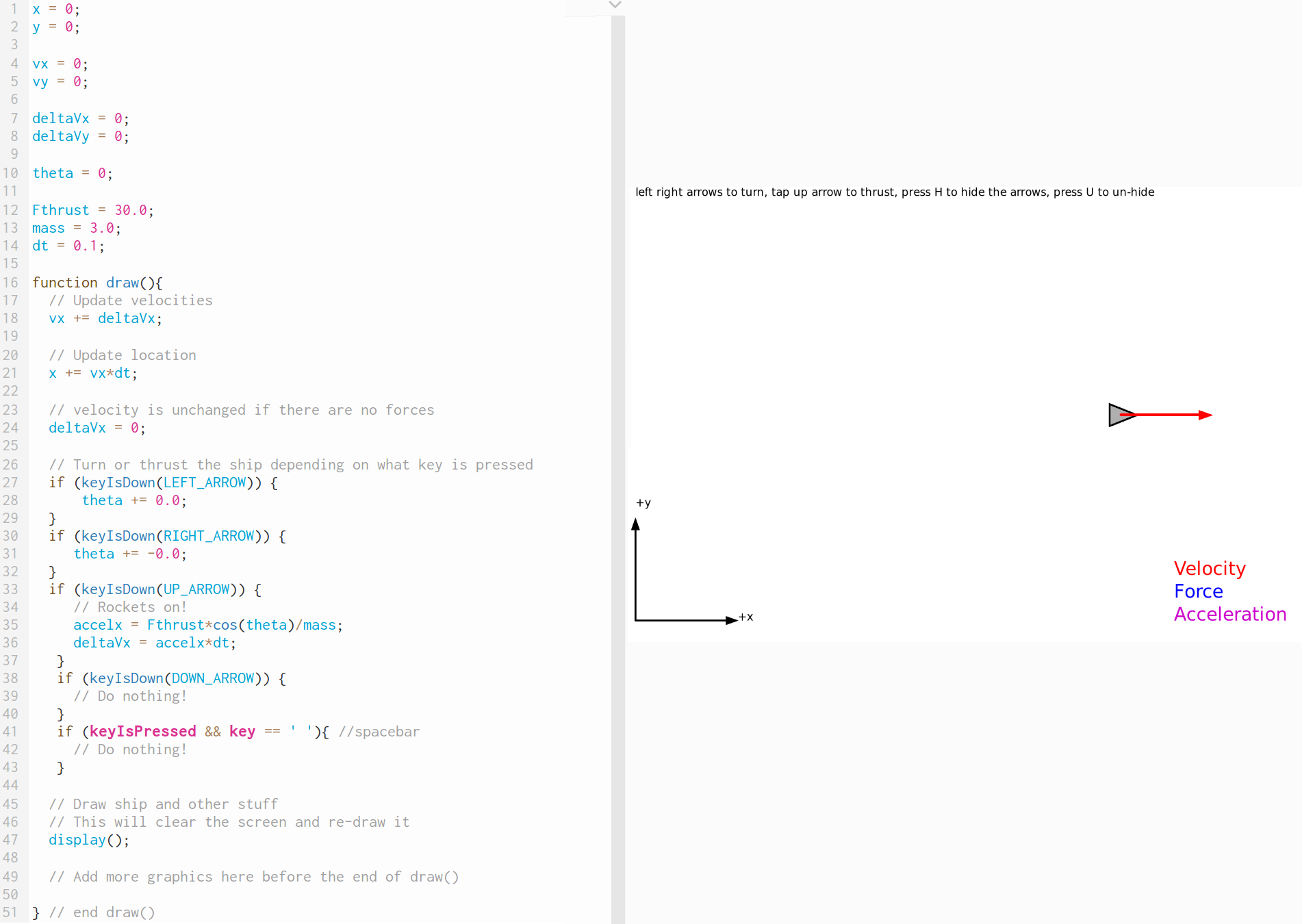}
    \caption{The code (left) and corresponding interactive (right) that the student sees at the beginning of the first exercise. This code is written with the processing javascript library p5.js. As a result, the code has a C/C++ like syntax except that draw() replaces main() and draw() is run 60 times per second until the user stops the program. The interactive (right) shows a ship traveling towards the right with constant velocity indicated by a red velocity vector. On the left panel the student sees about 50 lines of code, but about half of these lines are spaces or comments.}
    \label{fig:planetoids}
\end{figure*}

Fig.~\ref{fig:planetoids} shows what the student sees at the beginning of this exercise. Initially, the ship can only move in the $x$ direction and the first task is to allow the ship to rotate when the user presses the left and right arrow buttons by changing the value of $\theta$. It is worth commenting on Fig.~\ref{fig:planetoids} in detail because even at this stage there are a number of choices that have been made that could affect student learning. One important choice made to simplify the cognitive load for the student is to ``hide" a significant amount of code in the \texttt{display()} function. In this example, there are about 3$\times$ more lines of code defining the \texttt{display()} function than the $\approx 50$ lines of code that the student sees and modifies. 

Another important choice is to hide the variable types. There are no \texttt{float}, \texttt{int}, \texttt{double} or \texttt{var} declarations used to initialize the variables. Instead, variables are implicitly declared to be floating point decimals and the number of characters that the student sees is minimized. This syntax is essentially the same as used in matlab, which is a popular language for absolute beginner programmers. We use the processing javascript library p5.js for these exercises and as a result the code shown in Fig.~\ref{fig:planetoids} is javascript which does not produce an error for missing variable types. A possible drawback of postponing the discussion of variable types is that the difference between global and local variables is not explained at this stage. Students may not realize that \texttt{accelx}, which is only used and defined inside of an \texttt{if} statement, is a local variable while \texttt{deltaVx} is a global variable, but this is unlikely to cause a problem at this stage. Our philosophy is to explain subtleties like these in the step-by-step tutorial only if absolutely necessary for completing a particular exercise.

The structure of the program in Fig.~\ref{fig:planetoids} is an important choice that may affect student learning. The sections of the code are as follows:
\begin{enumerate}
\item Variable initializations 
\item the draw() function -- velocity and position advance
\item the draw() function -- keyboard inputs
\item the draw() function -- display() function followed by other user-defined graphics
\end{enumerate}
It is understood that the \texttt{draw()} function is run many times per second so that after the \texttt{display()} function is executed the program will go back to the beginning of \texttt{draw()} and advance the velocity and position again and go through the whole sequence again until the user presses stop. Because \texttt{draw()} is being run again and again, one could easily change the sequence so that, for example, the \texttt{display()} function would be first and the velocity and position advance would be last. The drawback of this approach is that when the student parses the code for the first time they would see the physics content of the code \emph{last}. Perhaps a computer science instructor would have written the exercise with that order since, from their point of view, there is nothing special about the physics part of the code.

Putting the physics section at the earliest place it can be shows the student how familiar ideas like $d = vt$ are implemented in a computer code before getting into unfamiliar syntax like keyboard commands. Although it is not explained to the student until much later in the energy exercise, the integration scheme is Euler-Cromer \cite{Cromer1981}. Following the physics section there is a line of code $\Delta v_x = 0$ which is accompanied by a comment ``velocity is unchanged if there are no forces". This is just a restatement of Newton's first law in a form that a computer can understand. Following this the program checks if the user is pressing certain buttons on the keyboard.

The drawback to this physics-first, keyboard commands later approach is that the student may not fully appreciate that the program holds on to the global variable \texttt{deltaVx}, which is determined from the keyboard command section, only using it again at the beginning of the \emph{next} iteration of \texttt{draw()}. In fact, many students struggle to appreciate that \texttt{draw()} is run again and again. An interesting line of inquiry would be whether this tacit understanding that the special function \texttt{draw()} is executed many times causes more confusion than it eliminates by keeping the code very compact.

In a couple of steps the user is asked to put non-zero values in the section of the keyboard input section that changes the angle of the ship. Then the student is asked to enable motion in the $y$ direction by imitating the code for advancing the velocity and position in the $x$ direction. Finally, the student is asked to determine the correct change in velocity due to a constant force (thrust) in the $y$ direction. This involves realizing that while $\cos \theta$ gives the component of the force oriented in the $x$ direction, one must use $\sin \theta$ to obtain the component of the force in the $y$ direction. In this way the activity assumes trigonometry knowledge.

At each step in the tutorial, the student can click links to see and interact with how the program should work at a particular stage, but without seeing the source code for the completed step. This is an important capability that gives the student instant guidance on whether they have completed a particular programming task correctly, leaving the instructor more time to spend on subtle issues.

Common mistakes that students make include forgetting to set $\Delta v_y = 0$, in which case the ship accelerates uncontrollably in the $y$ direction. Students also tend to do a quick copy paste of the acceleration code without changing the trigonometric function from cosine to sine. This causes $\Delta v_y = \Delta v_x$ and as a result the ship only travels on a diagonal line.

\subsection{Challenges}

Students must also implement 1-2 ``challenges". The challenges in this exercise include creating ``planetoids" (a word play on the astronomical term planetesimals) that drift across the screen using the \texttt{ellipse()} function and adding reverse thrusters when the down arrow is pressed (which can be done by copying the code from the up arrow and adding minus signs to change the direction of the force). Students can also allow the ship to shoot a projectile using the \texttt{point()} function and the code includes an \texttt{if} statement that detects if spacebar is pressed for this purpose. This task is the more difficult than the others because the projectile must be launched in the same direction as the ship whereas the planetoids can be given a random velocity using the \texttt{random()} function. One can also include the velocity of the ship when determining the velocity of the projectile as a fun illustration of Galilean invariance. Most students will just implement the reverse thrusters challenge.

For students who create ``planetoids" as their challenge, a interesting activity would be to ask the student to change the mass of the ship and the force of the rocket thruster to find the best combination for avoiding planetoids. The student may realize through this activity that it is only the ratio of the force to the mass that matters for determining the acceleration of the ship. Because of the length of Exercise 1, this is typically not done until Exercise 5, but with more class time this would be an interesting option.

\subsubsection{Student data}
\label{sec:survey}

\begin{figure*}
\begin{center}
    \includegraphics[width=3.4in]{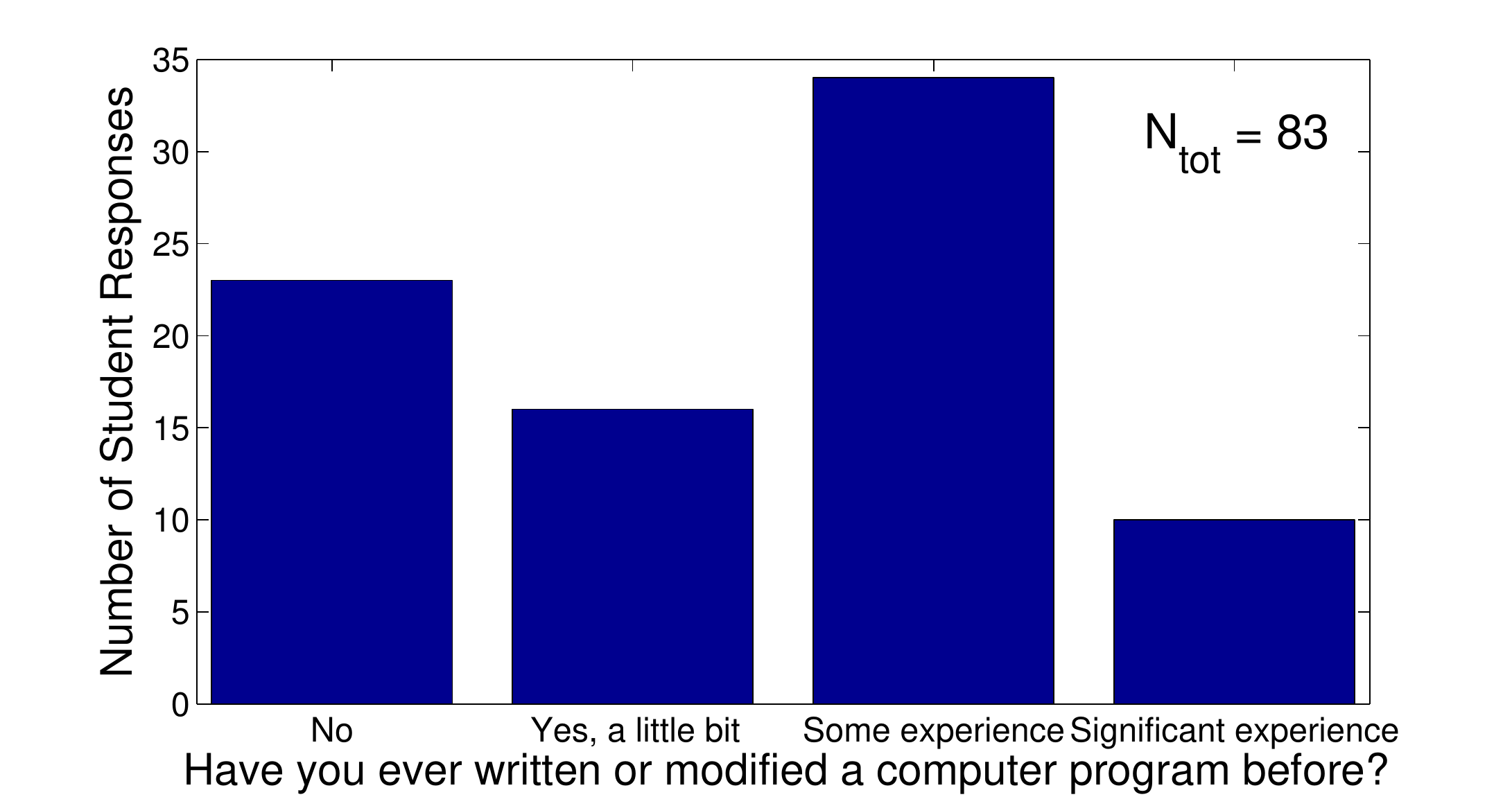}\includegraphics[width=3.4in]{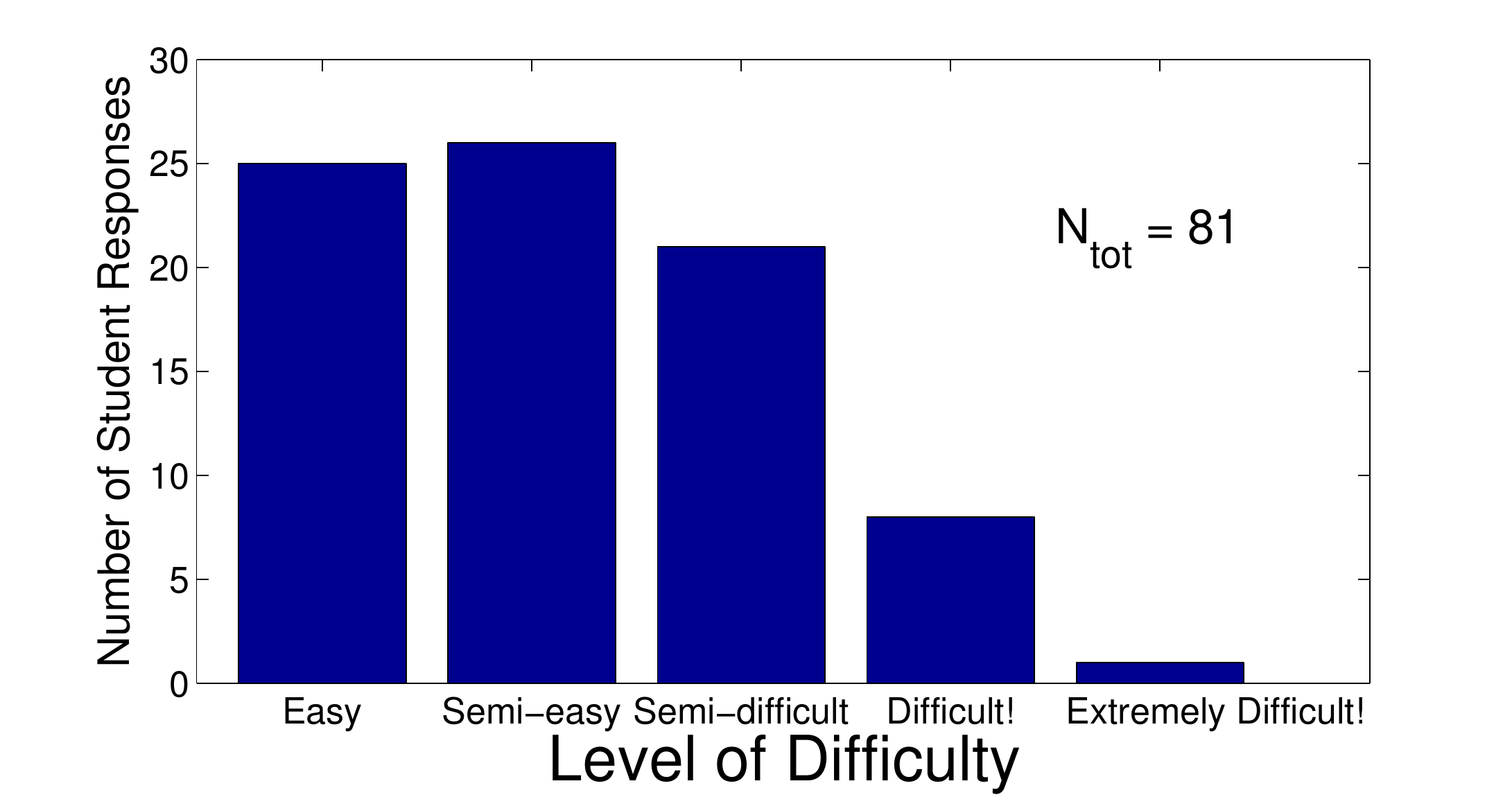}
    \includegraphics[width=3.4in]{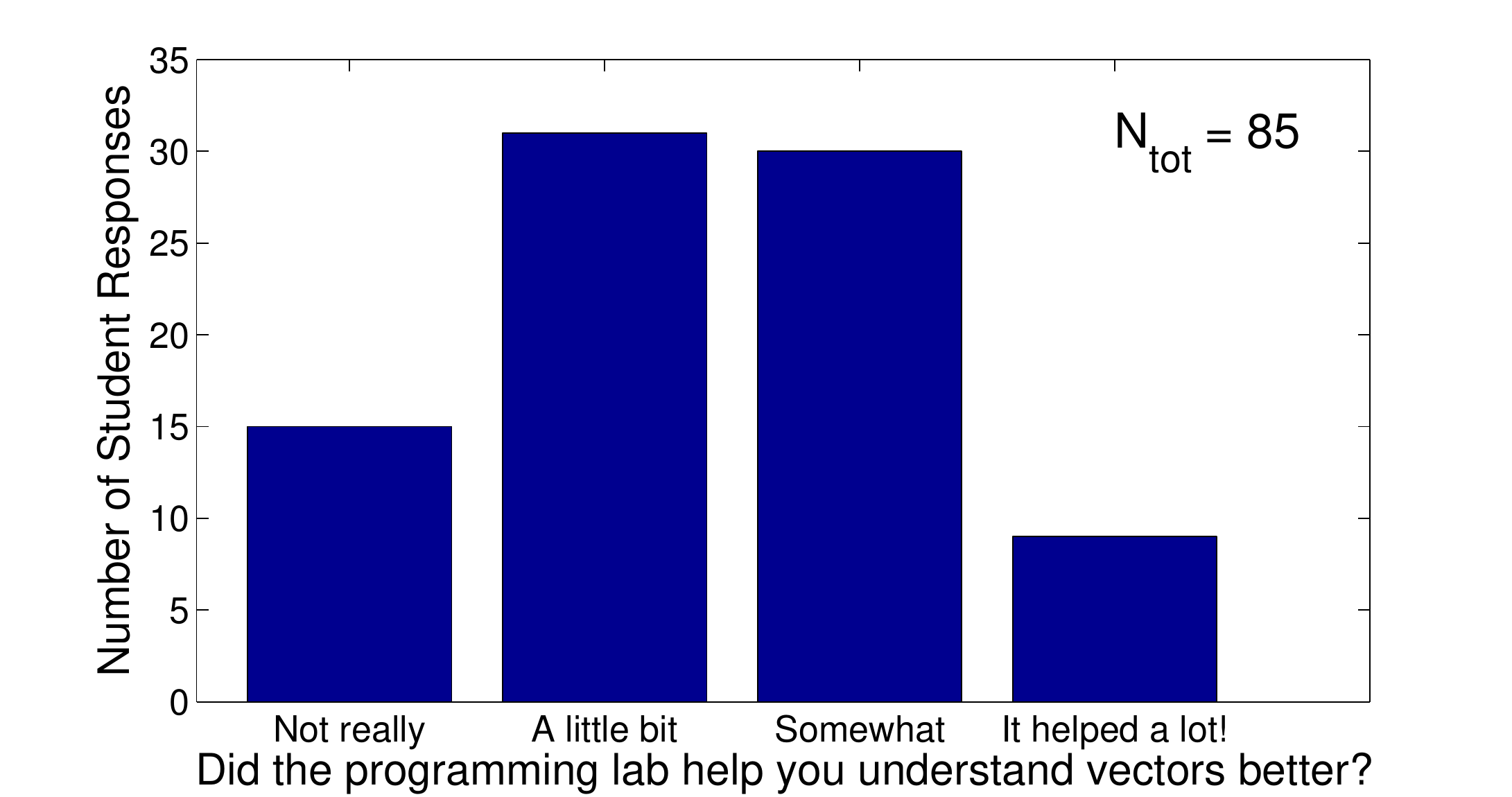}\includegraphics[width=3.4in]{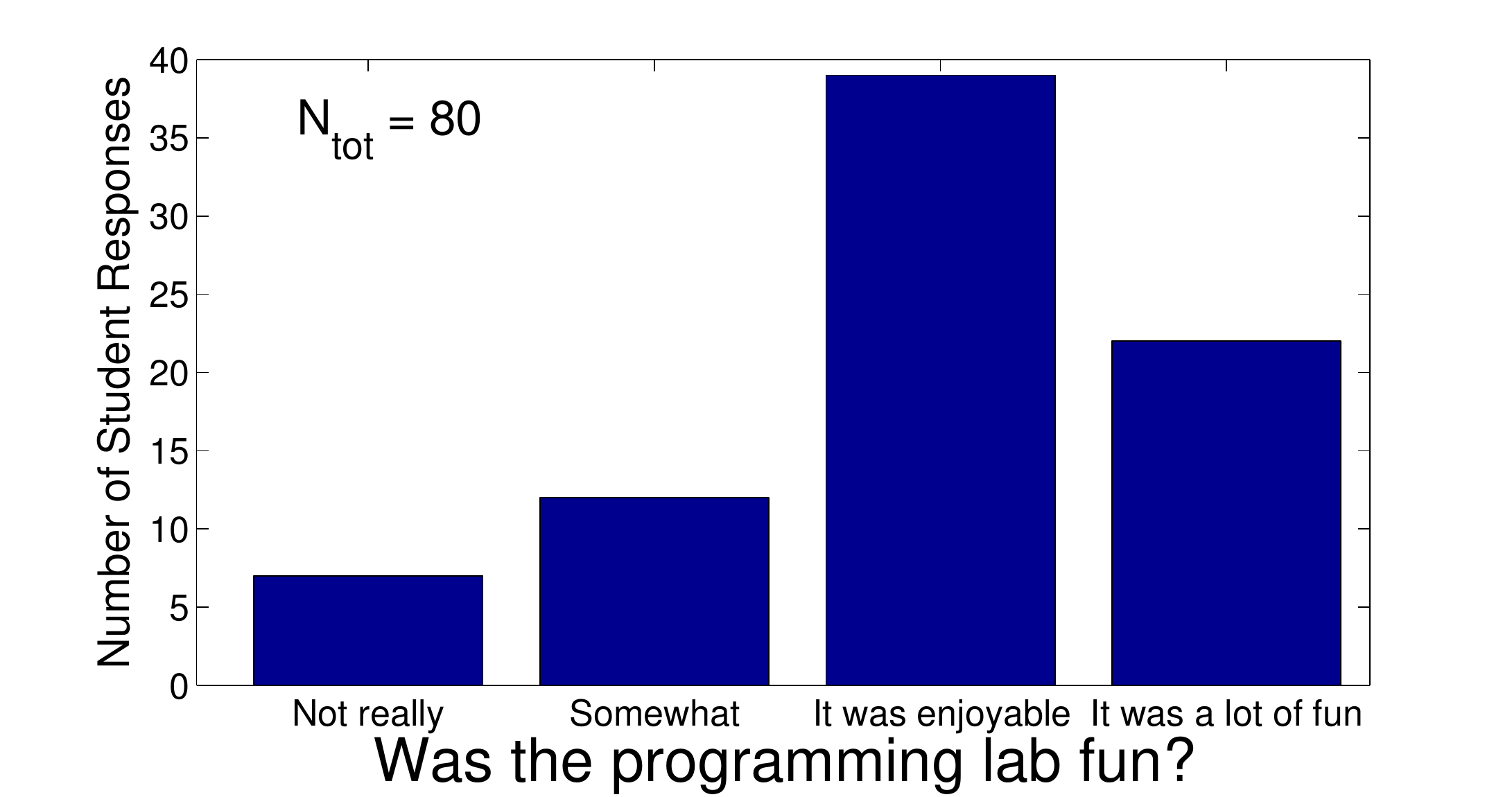}
    \end{center}
    \vspace{-0.6cm}
    \caption{Survey results from Ohio State Marion students who completed the first programming exercise (rocket in free space). Results are cumulative from four semesters of students (Spring 2015 - Fall 2016).} 
    \label{fig:survey}
\end{figure*}
%Porter: Different font sizes in x label of plots.

A few weeks into the course, after the student completes the first exercise there is a detailed online survey that probes their experience in completing the activity. While the questions in this survey are qualitative and involve student self-reporting, the results can offer insight on whether the level of difficulty of the first exercise is appropriate and whether students find the exercises to be enjoyable to complete. Appendix~\ref{ap:questions} shows the precise wording of the questions that were used in the survey. Fig.~\ref{fig:survey} summarizes the results of the survey from four semesters of students (Spring 2015 -- Fall 2016). The upper left plot in Fig.~\ref{fig:survey} shows that there is a significant number of absolute beginner programmers and weak programmers in the class. As noted in the appendix, the full wording of the ``some experience" option for this question is ``Yes, I have some experience with programming but I still feel like I have a lot to learn (For example: you took a programming course or are taking a programming course now, or have learned some programming on your own in some other context.)" The high percentage of students selecting this option is likely from students who take physics and an introductory C++ course concurrently. Many of these students may have been absolute beginner programmers at the beginning of the semester.

The upper right plot in Fig.~\ref{fig:survey} shows that the difficulty level seems to be appropriate for the population of students, with a significant number of students selecting ``Easy!". The lower right plot in Fig.~\ref{fig:survey} indicates that many of the students found the programming activities to be enjoyable or fun. The options in these plots correspond exactly to the options in the survey as discussed in the appendix. Students also have many positive things to say about the programming exercises in written evaluations at the end of the course after all of the exercises have been completed.

Regarding the bottom left plot in Fig.~\ref{fig:survey}, student reporting of learning gains should be met with skepticism. We regard the results for this question  "Did the programming lab help you understand vectors better?" primarily as a gauge of student perceptions of how connected (or disconnected) the programming activities are from other course material.
The mixed results from this question underscores the need to carefully integrate the exercise with instructional material and targeted assessment as we discuss later. In the absence of this, students may regard the activity as just another video game even when velocity and acceleration vectors are illustrated, as they are in this example. We outline our plan for more detailed assessment in Sec.~\ref{sec:assess}.
%Porter: I love the level of detail and the focus up to this point. From here I think the level of detail is entirely determined by where we want to publish.

\section{Exercise 2: Constant Gravitational Acceleration}

In Exercise 2, the student needs to add a variable $g$ to the beginning of the program and set it to the value 1.63 which is the gravitational acceleration constant on the moon in m$/$s$^2$ units. For simplicity, we follow processing's convention in using pixel values for distances which means that the 750x500 pixel window that shows the interactive is in reality 750 meters $\times$ 500 meters wide. In both the previous exercise and this exercise we use a time interval of $\Delta t = 0.1$ for each iteration of the draw() function. Although it is not explained to the student, this implies that a interactive running at 60 frames-per-second is really showing 6 simulated seconds per 1 second of actual time. In the end, these choices cause the ship to fall towards the ground over a period of several actual seconds, neither too slow nor too fast to be uninteresting.

Students are told to add gravity to the simulation by adding this line of code: \texttt{deltaVy += - g*dt} after the keyboard input section. This representation emphasizes gravity as a constant acceleration rather than as a force. This is another choice that simplifies the code but may increase the cognitive load in later exercises (particularly exercise 6 which includes a horizontal spring force and the force of the rocket engine).

The activity in this exercise is to try and land the ship on the bottom of the screen. Students are given an \texttt{if} statement to include in the code that allows the ship to land if the ship's height is close to the bottom of the screen and the velocity is small. The student also adds an \text{if} statement to trigger a ``Game over!" if the height of the ship becomes negative.

Students can customize the game by changing the variable $\theta$ so the ship initially points upward instead of to the right as in the previous exercise. Challenges include creating a time limit, or a limit to the amount of fuel.

\section{Exercise 3: Constant acceleration with an initial velocity}

\begin{figure}
    \centering
    \includegraphics[width=3.3in]{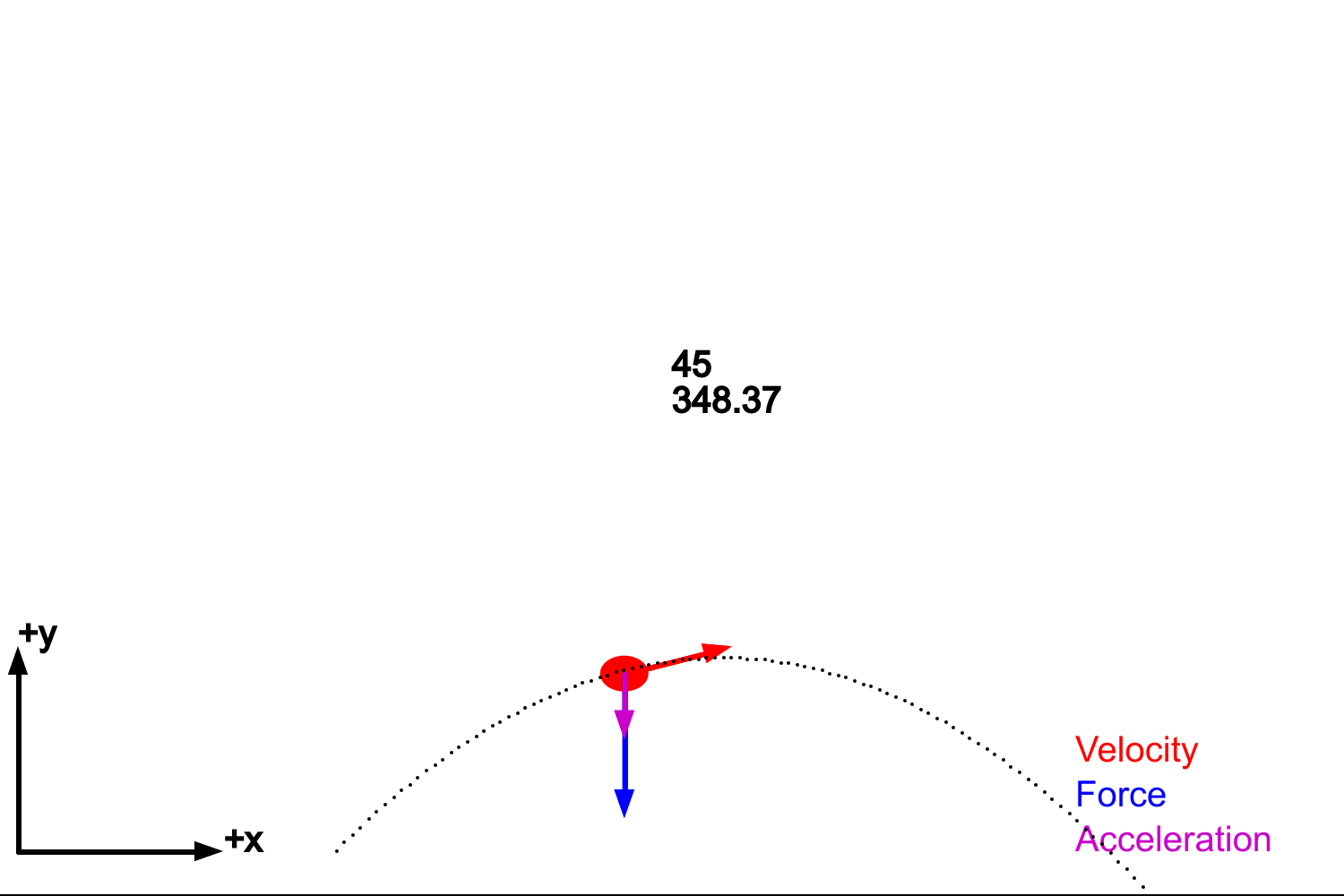}
    \caption{An interactive from the 3rd programming exercise. Students must configure the program to draw the expected trajectory before the projectile is launched. The keyboard controls the launch angle.}
    \label{fig:bellicose}
\end{figure}

In this exercise the student moves the position of the ship to the bottom left part of the screen and configures the if statement that checks if the spacebar is pressed so that the ship is given an initial velocity that depends on the angle that the ship is pointing. Students need to fill in the blank with the correct formula for $v_x$ and $v_y$ at the moment that the projectile is launched. Again, this draws from their knowledge of trigonometry to realise that $v_x = v_i \cos \theta$ and $v_y = v_i \sin \theta$. Even in a calculus based physics class, some students have difficulty arriving at this answer, perhaps, in part, due to the the unfamiliar environment of computer programming. One hopes that this task reinforces this concept when the student works on pen and paper calculations.

The next task, which is also the most difficult, is drawing the expected trajectory of the projectile before it is launched. This is an interesting goal because this trajectory changes depending on the launch angle of the projectile. Using step-by-step interactives the student can see and interact with the correct implementation of the trajectory in the program but they cannot  see the source code. Although the expected trajectory is just $y = y_i + v_{yi}t - (1/2)g t^2$ and $x = x_i + v_{xi} t$ students often struggle to produce these formulas and when they do they often use the current velocity $v_x$ and $v_y$ of the projectile in place of the initial velocities $v_{xi}$ and $v_{yi}$, which has the effect of making the trajectory change in strange ways after the projectile is launched. Once the students have completed this task they are asked to empirically show that a launch angle of 45$^o$ gives the furthest distance, a task that some students may have difficulty in proving analytically.

The challenge for this exercise is to configure the if statement that checks if the spacebar is pressed to only run ``then" code if the spacebar is pressed for the first time. Without this, it is easy to press the spacebar a moment too long and give the projectile its initial velocity again while it is already flying through the air. When this happens the projectile's motion and the expected trajectory do not agree to the extent that it does in Fig.~\ref{fig:bellicose}. So while this task is more computer science than physics, it serves to make the program more physically realistic. In this way this task supports our main goal of teaching students how to write computer programs to accurately solve physics problems.

\section{Exercise 4: Momentum and collisions}

In this exercise the student uses the ellipse() function to add a circular ``sticky blob" initially at rest that the ship will make a perfectly inelastic collision with. The student is given an if statement that detects if the $x$ position of the ship has passed closer than 1 radius away from the blob. The student must modify this if statement to properly detect a collision if both the $x$ and $y$ positions of the ship and the blob are close enough. Even in a calculus-based physics class, students have trouble using the Pythagorean theorem to determine the distance that separates the ship and the blob.

Another common problem encountered in this activity is that the students correctly determine the new velocity of the blob after the perfectly inelastic collision but they forget to set the velocity of the ship to this new velocity. In the next few timesteps the ship and blob, which are close to each other, will inelastically collide again and again until the blob reaches the same velocity of the ship. An important component of this activity is to carefully check that the final velocity of the ship and blob reach the values expected from momentum conservation. The problems students encounter are again more computer science than physics, but it reinforces the idea that the output of the code needs to be checked carefully to ensure that the program is modeling the physics correctly.

\section{Exercise 5: Torque}

\begin{figure}
    \centering
    \includegraphics[width=3.3in]{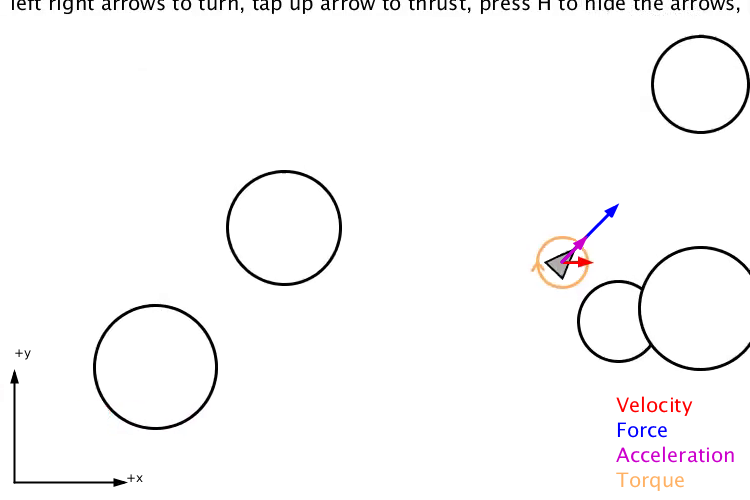}
    \caption{An interactive from the 5th programming exercise. Students must modify the mass, thrust and length of the ship to find a combination that gives the ship the most agility to avoid drifting ``planetoids". These variables affect both the acceleration in the direction of the ship and the angular acceleration when a constant torque is applied.}
    \label{fig:torque}
\end{figure}

In this exercise, the student modifies the code so ship spins at a constant rate unless thrusters apply a constant torque. The ship can still accelerate forward as before and no changes are needed to that part of the keyboard input section. Once the torque and rotation modifications are completed the students are asked to change the force of the thrusters, mass of the ship, and length of the ship to find optimal values that allow the ship to avoid hitting the planetoids. There is obviously no correct answer to this task, but it forces the student to think about how the acceleration and angular acceleration of the ship depend on these variables in a way that might not otherwise occur to them. There are no quantitative tasks in this activity, as in exercise 4.

A common mistake in this activity is when students forget to set the angular acceleration to zero at the beginning of the keyboard input section, which reflects the idea that the angular acceleration is zero if there is no torque on the system. When this happens the ship, once spinning, will rotate faster and faster. The students are given step-by-step interactives that clearly show this does not happen, but many students do not notice this.

For an interesting comparison to how an exercise like this might be put together in vpython, \cite{Buffler_etal2008} describes a similar programming activity where students configure a rocket to spin and thrust and are asked to predict the motion beforehand.

\section{Exercise 6: Spring forces and harmonic motion}

\begin{figure}
    \centering
    \includegraphics[width=3.3in]{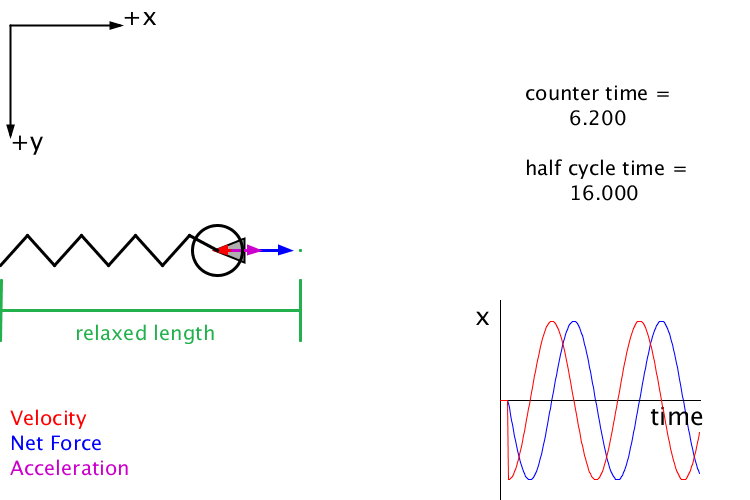}
    \vspace{-0.4cm}
    \caption{An interactive from the 6th programming exercise. The user can collide the ship with a ``blob" (i.e. the circle in the plot) that is attached to a spring, which begins periodic motion. Unlike other exercises, this interactive has a graph that charts the position and velocity of the blob with time.}
    \label{fig:spring}
\end{figure}

The final required exercise involves an interesting modification of Exercise 4 (momentum) where the sticky blob is attached to a spring. Initially the spring is drawn on the screen (as shown in Fig.~\ref{fig:spring}) and as before the blob is initially at rest, but there is no restoring force implemented in the code yet and the ship can fly into the blob and collide with a perfectly inelastic collision and carry the blob off the screen. Students must modify the code to add the restoring force of the spring so that when the ship collides with the blob, this begins harmonic motion. As seen in Fig.~\ref{fig:spring}, a graphing system is used here to plot the position and velocity of the ship as this motion occurs. The code for this graphing system is hidden away in the \texttt{display()} function.

An interesting aspect of this activity is that the ship is ``stuck" to the blob but user can still fire the thrusters to increase or decrease the amplitude of oscillations. The interactive activity is for the students to show that the ``half-cycle time" does not depend on the amplitude of the motion. The program includes a simple counter of how long it takes the blob to go from the relaxed position to maximum compression or extension and back again. Even with a simple scheme like Euler-Cromer integration \cite{Cromer1981}, one can easily show with good accuracy that the period and amplitude are independent of each other (\cite{Cromer1981} argues that the scheme performs rather well for oscillatory problems). The tutorial only refers to the ``half-cycle time" in an intentional effort to make students think about the motion and the code in order to realize that this is just half of the period. Students must calculate the expected ``half-cycle time" from the spring constant and the mass of the blob and ship with the only hint being $\omega = \sqrt{k/m}$.

The challenge in this exercise is to add linear damping ($b \neq 0$) to the system and measure the half-cycle time in that case, comparing it to the expected half-cycle time from $\omega = \sqrt{k/m - (b/2m)^2}$.  If $b$ is chosen to be neither too big (overdamped -- less than one oscillation) nor too small (almost no damping) one can obtain good agreement between the expectation and the measured period in the program. 

It should be said that at any point in the execution of the program, the student can fire the rocket thrusters to try and push the blob left or right. For example, the student might try to fire the thrusters in a futile but entertaining effort to try to resist the restoring force and any damping.

An additional virtue of the spring exercise (and one reason for the ubiquity of spring systems in computational physics tutorials) is that instructors often mention to students that drag forces in real physical systems are more often proportional to the velocity squared. Students could easily modify the program to include this more realistic drag force if desired. But in a typical course without numerical exercises, an instructor would have to say that, due to the complexity of the differential equation in this case, there is no method at the students' disposal that would allow them to make any progress. Many authors have argued that situations like these -- systems that are analytically frustrating or intractable -- are an important motivator for including numerical exercises in introductory physics \cite[e.g.][]{muppet1993,Buffler_etal2008,Chabay_Sherwood2008}.

\section{Exercise 7: Energy conservation}

\begin{figure}
    \centering
        \includegraphics[width=3.3in]{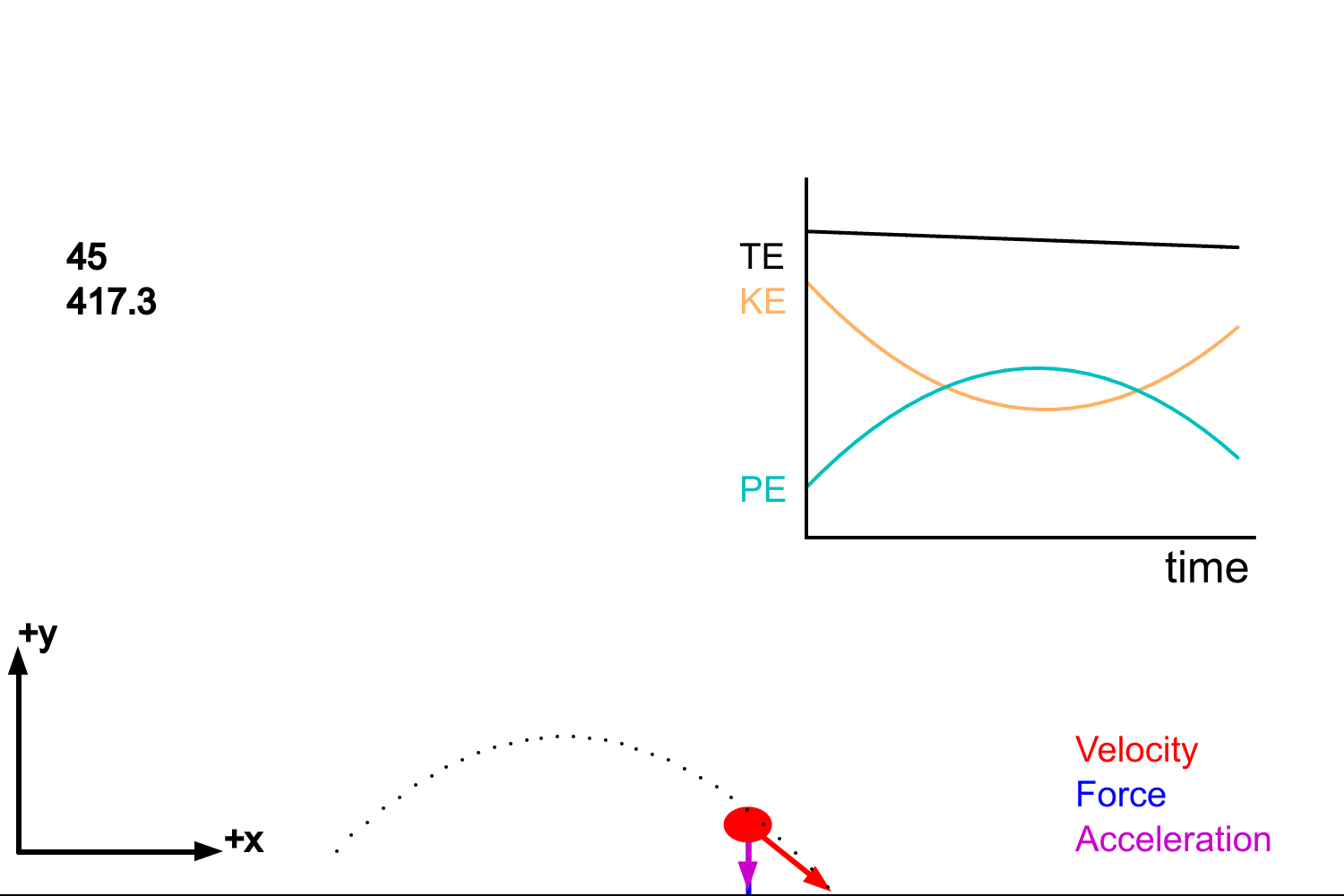}
    \caption{An interactive from an optional extra credit exercise that examines total energy conservation in the 3rd exercise that resembles the popular game ``angry birds". The graph shows that when the kinetic energy and potential energy are added together the result (black line) is not perfectly constant as expected. The exercise discusses this result and guides the student to fix this by modifying part of the code.}
    \label{fig:energy}
\end{figure}

This exercise adds a graph to Exercise 3 (``bellicose birds") and students notice that when the kinetic energy and potential energy of the projectile are added together one does not quite get perfect energy conservation as one would expect, regardless of the initial launch angle of the projectile. In Fig.~\ref{fig:energy}, if one looks closely, one can see a slight downward slant to the black line that shows the total energy instead of a horizontal line as one would expect.  Because the integration scheme uses Euler-Cromer integration \cite{Cromer1981}, the total energy is not perfectly conserved and each time step loses a small amount of energy. If the student allows the projectile to ``fall through the floor" (as they are instructed to do in the exercise by commenting out certain lines of code) this slant becomes much more obvious. Philosophically this is an important moment because it reveals that the computer did not determine the trajectory of the projectile perfectly. Instead, students will have to change something about the program to make this happen.

Students are asked to examine if the velocity update is to blame for the energy conservation problem by modifying the code to plot $v_y$ vs $t$ to check if this is a diagonal line with negative slope as expected. Because the program updates the velocity with $\Delta v_y = a_y \Delta t$ this turns out to look fine and the next part of the tutorial instructs the student that the problem must be in the position update. If one looks closely at the code in Fig.~\ref{fig:planetoids}, one notices that the $x$ position is updated with the \emph{already updated} velocity, which is the Euler-Cromer method. In Exercise 1, the students mimic this same code to update $v_y$ and the $y$ position.

As is well known, changing the position update to use the half-advanced velocity ($v_y + \Delta v_y / 2$), one can change the integration scheme from Euler to trapezodial (a.k.a. "midpoint") and perfect energy conservation is restored for constant acceleration problems.  This can be explained to the students in terms of calculating how far a car will travel when it is accelerating at a constant rate from zero to 60~mph. If we tried to model this in our program using one very large time step, say $\Delta t = 5$~seconds, then the original program would say that the distance traveled is $ v_f \Delta t$ because the original program uses the updated velocity to calculate the distance traveled. But this would be like assuming that the car traveled 60~mph for a full five seconds, which would be too far.  If we switch the order of the position advance and the velocity advance so we use the un-updated velocity ($v_i =  0$~mph) times the time interval to determine the distance traveled we will get zero for the distance. The right answer is to use the average velocity over the interval that gives the correct distance traveled. We can implement this in the program by using the half-advanced velocity in the position advance.

This is the newest exercise and very few students have completed it so far. Student experiences with this exercise will be discussed in future work. We include it here because it illustrates how topics like Euler vs Trapezoidal integration can be discussed at an introductory (i.e. pre-calculus) level. The exercise is also a stepping-stone towards simulating gravitational problems which require a better-than-Euler scheme to convincingly produce elliptical orbits and Kepler's laws. Discussing the Trapezoidal method for the first time in such an exercise could be more confusing to the student than introducing it in a constant gravitational acceleration case.

\section{Plans for Assessment}
\label{sec:assess}

The Force Concept Inventory (FCI) \cite{FCI} is a well-studied metric that is often used to quantify student learning and learning gains. What is less well-known is that M. Dancy created an animated version of the FCI, using essentially the same java-based framework as physlet physics \cite{physlet}, and compared results from a significant number of students completing the animated version and the often-used written version \cite{Dancy2006}. From student interviews \cite{Dancy2006} found that the animated version gave students a more accurate understanding of the questions being asked. We comment here to mention that many of the questions that did not have any diagrams in the written version have animations depicting each multiple-choice option where originally there was just a sentence describing the motion. Thus it is understandable why the animated version seemed to do a better job of communicating the questions to the students. We view the animations as a substantial improvement over the written FCI, and it seems like a natural place to start in assessing whether students gain a better conceptual understanding of physics from working through the programming exercises.

In particular, because of our emphasis on a ship traveling in free space, questions 21 through 24 on the FCI seem particularly relevant. In these questions, students are asked to identify the correct trajectory of a rocket drifting through free space, followed by a period of constant thrust. With questions 21 \& 22, there are diagrams in the written version that the animated version emulates. With questions 23 \& 24, which describe the motion of the ship after the thrust has turned off, in our opinion the animated version does a significantly better job of explaining the nature of the question.

We re-created the animated FCI using files from the supplemental info for \cite{Dancy2006}, and using libraries from the CD that comes with the physlet physics textbook \cite{physlet}. To create a version of the animated FCI that does not require the student's computer to run java applets (which are increasingly obsolete due to security risks), we had to make screen capture videos from a computer with a java-capable browser, and incorporate these screen capture videos (which involve commonly-used mp4 file formats) into an online quiz. After some work we were able to re-create all 30 animated FCI questions in this way.

%We only have one semester of data from students taking the animated FCI at the beginning of the course and taking questions 21-24 again shortly after the submission of the first programming exercise, and only about 12 students of about 40 completed both tasks. The data for questions 21-24 for these students do not show any significant learning gains. One can speculate that this may be partly explained by the lack of resemblance between the cartoon image of a ship used in the animated FCI and the triangle-shaped ship we use in our programming exercises.

In the future, we plan to increase this sample size substantially, and we plan to integrate animated FCI questions into a custom-built website (stemcoding.osu.edu) where students will complete the exercises using an in-browser editor with bug-finding capabilities. This website will be configured to ask students animated questions as a kind of "pre-lab" immediately before they start working on the code and immediately after they submit their code for grading, so as to avoid a situation where student performance could be attributed to something besides the programming exercises. This website, which can be thought of as a p5.js-based learning management system, also has a capability for teachers to quickly view submissions and give feedback to students in a much more streamlined way than would otherwise be possible. This is an important capability for any effort to scale up these kinds of programming activities to significantly larger classes.  This website was used for the first time in spring 2017 physics courses at OSU's Marion campus.

%Regarding the concern about the appearance of the ship, we may modify the animated FCI questions 21-24 to more so resemble the ship used in the programming exercises, and significantly increase the frame rate of these animations from about 2 frames per second (as originally designed) to closer to 60 frames per second, which may help students see the connection between the questions and the programming exercises.

%Discussion of Animated FCI and how we brought it back from the dead.

%\section{Literature Review}

%{\bf This is not actually a section. It is a stand in. These references should be worked into the introduction and then the literature review section should be eliminated.}

%Another important paper is \cite{Kohlmyer_etal2009} but it is more of a test of the entire Matter \& Interactions framework, not just the programming labs.

%There is also an extensive review paper \cite{review2014} where programming exercises are discussed briefly.

%Finally, there is an AJP issue dedicated to programming exercises incorporated in physics curricula (Vol. 76, year 2008).

\section{Summary and Conclusion}

We present a comprehensive set of programming exercises for introductory physics that are appropriate for algebra-based and calculus-based physics classes at either the high school or college level where students may be absolute beginner programmers. Creating content at this level is a challenging task and we argue that there are a number of important considerations: (1) the code that the student sees must be well commented and involve 75 or fewer lines of code, (2) the activities must be easy-to-use with browser-based coding tools, (3) these exercises must emphasize interactivity and include goal-oriented game-like activities, (4) the directions must be step-by-step with the ability to interact with intermediate stages of the "correct" program, and (5) the exercises must be thoughtfully integrated into the physics curriculum.

In an introductory class at OSU where a substantial fraction of the students are weak or absolute beginner programmers, student survey data ($N \approx 80-85$) seems to confirm that the first exercise, which produces an interactive simulation that resembles the classic game ``asteroids", is at an appropriate difficulty level, and that students overall seem to enjoy this exercise. 

We describe five other required exercises and one extra credit exercise. Following the first exercise, students complete an exercise where gravity is added to the simulation by adding two lines of code. Another exercise, which is not unlike the popular game angry birds, asks students to configure the code to draw a line of the expected trajectory of a projectile launched from the ground and to confirm that the max distance occurs for a 45$^o$ launch angle.  Exercises following this return to the ship in space theme and sequentially build upon another, discussing momentum, torque and harmonic motion. Game-like goal-oriented activities are included throughout, such as an activity to determine the best values of the ship's mass, length and force of the thrusters to navigate through an asteroid field. In another activity students use the ship's thrusters to increase the amplitude of a spring system and demonstrate that the period of motion is insensitive to the amplitude. Finally, an extra credit exercise takes the angry-birds-like code and reveals that the total energy is not perfectly conserved. A gentle introduction to the Euler-Cromer and trapezoidal (a.k.a. midpoint) methods of determining the path of motion is presented that does not assume calculus knowledge. 

Like other authors \cite[e.g.][]{Chabay_Sherwood2008}, our ultimate goal is to improve the student's conceptual knowledge of physics and ability to solve problems. We outline a plan for detailed assessment of student learning using an animated version of the Force Concept Inventory \cite{Dancy2006} that we have reproduced from various sources to allow the animations to play on modern computers. These questions will be integrated into a custom-designed p5.js learning management system in order to query students shortly before they begin and after they finish a programming exercise. This p5.js learning management system is in operation and it will allow students in physics classes at OSU's Marion campus starting in spring 2017 to complete the exercises using an in-browser code editor and it will give teachers a convenient interface for the otherwise tedious task of providing student feedback and grades. We welcome inquiries from collaborators who may wish to use these animations in their high school or early college-level courses, or who may be interested in using the p5.js learning management system.

The full set of exercises and code described here is available at \url{http://compadre.org/PICUP} 

\acknowledgements

The authors thank Kyle Decot and Michael Hardesty for their collaboration on a p5.js learning management system. CO thanks Annika Peter, Gregory Ngirmang, and Kelly Roos for discussions. This project was made possible through a Connect and Collaborate Grant, a program supporting innovative and scholarly engagement programs that leverage academic excellence of The Ohio State University in mutually beneficial ways with external partners.

\appendix
\onecolumngrid

\section{Survey Questions}
\label{ap:questions}

As discussed in Sec.~\ref{sec:survey}, a few weeks into the course and after the student submits the first programming exercise for grading/feedback, the students are asked a series of questions in an online form. Fig.~\ref{fig:survey} shows the results from four semesters of students taking this survey. The precise questions are as follows:

\subsection{Question \#1}

Question: Have you ever written or modified a computer program before?

Option A: No, I have never done any programming before

Option B: Yes, I have done a little bit of programming before

Option C: Yes, I have some experience with programming but I still feel like I have a lot to learn (For example: you took a programming course or are taking a programming course now, or have learned some programming on your own or in another context)

Option D: Yes, I have a significant amount of programming experience (ex. completed more than one programming course, or significant programming experience outside of class or in projects).

Option E: Prefer not to answer (but still get the point for the question)

\subsection{Question \#2}

Question: From the point you got the planetoids program working until you finished the programming lab (including one of the challenges), {\bf compared to the difficulty of completing other Physics 1250 labs}, how difficult was it to complete the programming tasks described on the planetoids page?

Option A: Easy!

Option B: Semi-Easy

Option C: Semi-difficult

Option D: Difficult!

Option E: Extremely Difficult!

Option F: Prefer not to answer (but still get the point)

\subsection{Question \#3}

Question: Did the programming lab help you understand vectors better?

Option A: Not really

Option B: A little bit

Option C: Somewhat

Option D: It actually helped a lot

Option E: Prefer not to answer (but still get the point)

\subsection{Question \#4}

Question: Was the programming lab fun?

Option A: Not really

Option B: Somewhat

Option C: It was enjoyable

Option D: Actually it was a lot of fun

Option E: Prefer not to answer (but still get the point)

\bibliographystyle{apsrev}
\bibliography{main}
\end{document}